\documentclass[aps,prd,nofootinbib,showpacs,
tightenlines,superscriptaddress,preprint]{revtex4}

\usepackage{color}

\def\d{{\rm d}}
\usepackage{amsfonts}
\begin{document}

\title{Quantum Gowdy $T^3$ Model:
Schr\"odinger Representation with Unitary Dynamics}
\author{Alejandro Corichi}\email{corichi@matmor.unam.mx}
\affiliation{Instituto de Matem\'aticas, Universidad
Nacional Aut\'onoma de M\'exico, UNAM-Campus Morelia,
A. Postal 61-3, Morelia, Michoac\'an 58090, Mexico.}
\affiliation{Institute for Gravitation and the Cosmos,
Pennsylvania State
University, University Park PA 16802, USA.}
\author{Jer\'onimo Cortez}\email{jacq@fciencias.unam.mx}
\affiliation{Departamento de F\'\i sica,
Facultad de Ciencias, Universidad Nacional Aut\'onoma de M\'exico,
A. Postal 50-542, M\'exico D.F. 04510, Mexico.}
\author{Guillermo A. Mena Marug\'an}\email{mena@iem.cfmac.csic.es}
\affiliation{Instituto de Estructura de la Materia,
CSIC, Serrano 121, 28006 Madrid, Spain.}
\author{Jos\'e M. Velhinho}\email{jvelhi@ubi.pt}
\affiliation{Departamento de F\'{\i}sica, Universidade
da Beira Interior, R. Marqu\^es D'\'Avila e Bolama,
6201-001 Covilh\~a, Portugal.}

\begin{abstract}
The linearly polarized Gowdy $T^3$ model is
paradigmatic for studying technical and conceptual
issues in the quest for a quantum theory of gravity
since, after a suitable and almost complete gauge
fixing, it becomes an exactly soluble midisuperspace
model. Recently, a new quantization of the model,
possessing desired features such as a unitary
implementation of the gauge group and of the time
evolution, has been put forward and proven to be
essentially unique.  An appropriate setting for making
contact with other approaches to canonical quantum
gravity is provided by the Schr\"odinger
representation, where states are functionals on the
configuration space of the theory. Here we construct
this functional description, analyze the time
evolution in this context and show that it is also
unitary when restricted to physical states, i.e.
states which are solutions to the remaining constraint
of the theory.
\end{abstract}

\maketitle

\section{Introduction}

In the quest for a quantum theory of gravity, the use
of simple models has proven to be very effective. The
simplest possible models, where the most symmetries
are imposed from the outset \cite{misner1}, have
become important for the study of Planck scale
modifications to the Big Bang scenario (see, e.g.,
Ref. \cite{lqc}). However, these models suffer from an
oversimplification since all inhomogeneous degrees of
freedom are neglected. A natural question is how the
inclusion of these inhomogeneous modes affects the
qualitative picture near the singularity that the
homogeneous models possess. In this regard, the
linearly polarized Gowdy $T^3$ model is a natural
candidate for a detailed study. It is the simplest
inhomogeneous, spatially closed, cosmological model in
vacuo \cite{gowdy}. One important reason for the
appeal of such a model is that, after a convenient
almost complete gauge fixing and the introduction of a
geometrically motivated internal time, the model
becomes soluble. Any solution of the full set of
Einstein equations can be obtained from the solutions
of an auxiliary scalar field in a fixed fiducial
background. However, this auxiliary scalar field
system is not unique. Different ``field
parametrizations'' of the metric may give rise to
different scalar field systems. Classically they are
all equivalent, but in the quantum theory this may not
be so. In addition, the quantization of field systems
possesses an infinite degree of ambiguity, even if one
restricts all considerations to standard
quantizations, e.g., of the Fock type. As a
consequence, there exist in principle infinitely many
inequivalent quantizations of the Gowdy $T^3$
midisuperspace model.

Among the different possibilities available in this
route to quantization, two field parametrizations have
received special attention in past years. One of them
can be considered a somewhat conventional field
parametrization from the viewpoint of a dimensional
reduction of the model \cite{pierri}. However, this
proposal for the choice of fundamental field has the
undesirable property of not implementing the dynamics
(generated by the internal notion of time) unitarily.
Actually, although this lack of unitarity was first
proven \cite{non-uni} for a ``natural quantization''
of the associated scalar field, introduced by Pierri
\cite{pierri}, it has been recently shown that there
exists no Fock quantization with a unitary dynamics,
at least if one also demands an invariant unitary
implementation of the gauge group that remains on the
model after gauge fixing \cite{cmv}. To solve this
problem, a new field parametrization, together with an
essentially unique quantum representation, was
recently introduced. In this case, not only the
evolution is unitary and the gauge group is naturally
implemented, but it has been shown that any other Fock
quantization of the new field with such properties is
unitarily equivalent to the constructed one
\cite{cmv,ccm1,ccm2,ccmv}. Furthermore, the adopted
field parametrization turns out to be unique in a
precise sense under the condition of the existence of
a Fock representation (FR) with an invariant unitary
action of the gauge group and a unitary dynamics
\cite{cmv}. These results were mainly formulated in
the language of Fock space, which is natural from the
perspective of a scalar field in a fixed background.

On the other hand, quantum gravity in its canonical
formulation is commonly defined in the Schr\"odinger
functional picture, where states are functionals on
the configuration space of the theory. Therefore, it
is important to have a Schr\"odinger functional
description of any symmetry reduced model, such as the
Gowdy $T^3$ model. The purpose of this paper is to
present this description for the quantization which
admits a unitary time evolution \cite{ccm1,ccm2}, and
analyze the implementation of such a unitary evolution
in this framework, both before and after imposing the
remaining constraint of the theory.

We will adopt here the same viewpoint as in Refs.
\cite{ccm1,ccm2}: instead of working with a fixed
quantum representation and considering the unitary
implementability of the family of symplectic
transformations defined by the evolution (together
with the corresponding unitary evolution operator), we
will construct the associated 1-parameter family of
representations. Notice that this is precisely the
family of representations which is obtained by
``evolving in time'' a fixed GNS state, and hence the
complex structure defining the FR. The equivalence
between the two viewpoints is then established by the
fact that evolution between any two given times admits
a unitary implementation if, and only if, the
corresponding representations are unitarily
equivalent.

Finally, we note that the 1-parameter family of
complex structures that gives rise to the 1-parameter
family of unitarily equivalent representations can be
obtained both on the canonical phase space (the space
of Cauchy data for the auxiliary scalar field) or on
the covariant phase space (the space of solutions).
Since we are interested in the canonical functional
description, we will obtain the family of complex
structures directly on the canonical phase space. As a
particular consequence of unitarity, we will obtain a
family of mutually equivalent Gaussian measures in the
(quantum) configuration space.

The structure of the paper is the following. In Sec.
\ref{ccmv-quant} we recall the quantization of the
linearly polarized Gowdy $T^3$ model constructed by
Corichi, Cortez and  Mena Marug\'an, in which the time
evolution is implemented unitarily \cite{ccm1,ccm2}.
In Sec. \ref{sec:3} we construct the Schr\"odinger
representation (SR) corresponding to this (unique)
Fock quantization. In Sec. \ref{sec:5}, we implement
the canonical notion of time evolution within the
Schr\"odinger description, showing explicitly the
equivalence of the family of representations at
different times. The conclusions are presented in Sec.
\ref{sec:6}.

\section{The quantum Gowdy model}
\label{ccmv-quant}

In this section we will review the quantization of the
Gowdy $T^3$ cosmological model as performed in Refs.
\cite{ccm1,ccm2}. We will start with a description of
the classical model and its dynamics.

\subsection{The classical model}
\label{ccmv-quant-1} The linearly polarized Gowdy
$T^3$ model describes globally hyperbolic
four-dimensional vacuum spacetimes, with two commuting
hypersurface orthogonal spacelike Killing fields and
compact spacelike hypersurfaces homeomorphic to a
three-torus. In a coordinate system
$\{(t,\theta,\nu,\delta), \, t\in \mathbb{R}^{+}; \,
\theta,\nu,\delta \in S^1\}$ with $(\partial_{\nu})^A$
and $(\partial_{\delta})^A$ being the hypersurface
orthogonal Killing fields, the line element can be
expressed as
\begin{equation}
\label{line-element} \d s^{2}=e^{\gamma
-(\xi/\sqrt{pt})-\xi^{2}/(4pt)} \left( -\d
t^2+\d\theta^2\right)+e^{-\xi/\sqrt{pt}}
t^{2}p^{2}\d\nu^2+e^{\xi/\sqrt{pt}}\d\delta^2
\end{equation}
after a gauge fixing procedure which removes all the
gauge degrees of freedom except for a homogeneous one
\cite{ccm2}. The spatially homogeneous variable $p$ is
a positive constant of motion. On the other hand, the
fields $\xi$ and $\gamma$ depend only on the time
coordinate $t$ and the spatial coordinate $\theta$.
The field $\gamma$ is completely determined by $\xi$,
by $P:=\ln p$ and by their respective momentum and
configuration (canonically) conjugate variables,
$P_{\xi}$ and $Q$ (see Ref. \cite{cmv} for details).
Therefore, all local degrees of freedom reside in the
field $\xi$.

As we have mentioned, the model is just partially
gauge fixed: there is still a global constraint,
\begin{equation}
\label{homo-const}
{\cal{C}}_{0}=\frac{1}{\sqrt{2\pi}}\oint \d\theta
P_{\xi}\xi'=0,
\end{equation}
which comes from the homogeneous part of the
$\theta$-momentum constraint. Here, the prime denotes
the derivative with respect to $\theta$.

After the reduction process, the Hamiltonian
becomes\footnote{We set $4G/\pi=c=1$, $G$ and $c$
being Newton's constant and the speed of light,
respectively.}
\begin{equation}
\label{hamil} H = \frac{1}{2}\oint \d\theta
\left(P_{\xi}^{2}
+(\xi')^{2}+\frac{1}{4t^{2}}\xi^{2}\right).
\end{equation}
Note that, since the reduced Hamiltonian does not
depend on the degrees of freedom $Q$ and $P$, these
are constants of motion, and will be obviated in our
subsequent discussion.

Thus, the resulting system consists of a real scalar
field $\xi$ subject to the constraint
(\ref{homo-const}). Its Hamiltonian (\ref{hamil}) is
that of a massless field with a quadratic time
dependent potential $V(\xi)=\xi^{2}/(4t^{2})$
propagating in a (fictitious) background
$({\cal{M}}^{(f)},g_{AB})$, where
${\cal{M}}^{(f)}\simeq S^{1}\times \mathbb{R}^{+}$ and
$g_{AB}=-(\d t)_{A}(\d
t)_{B}+(\d\theta)_{A}(\d\theta)_{B}$.

We will now describe the (linear) dynamics of this
field system, starting with the covariant description.
The reduced Hamiltonian (\ref{hamil}) leads to the
field equations
\begin{equation}
\dot{\xi}=P_{\xi},\quad\quad
\dot{P}_{\xi}=\xi''-\frac{\xi}{4t^{2}},
\label{hamiltonianeqs}\end{equation} where the dot
denotes the derivative with respect to $t$. Hence, the
field $\xi$ satisfies the second order differential
equation
\begin{equation}
\label{field-eq}
\ddot{\xi}-\xi''+\frac{\xi}{4t^{2}}=0.
\end{equation}
Since the general solution is most conveniently
expressed in Fourier series, let us introduce the
notation
\begin{equation}e_{k}:=\frac{e^{-ik\theta}}{\sqrt{2\pi}}
\quad \quad \forall k \in \mathbb{Z}.
\end{equation}
With respect to some reference (``initial'') time
$t=t_{0}$, all smooth solutions can then be written as
\cite{ccm2}
\begin{equation}
\label{solutions}
\xi(t,\theta)=\sqrt{t}\left[q_{0}+p_{0}\ln(t)\right]
+\sum_{k\in
\mathbb{Z}-\{0\}}\left[b_{k}(t_{0})\,G_{k}^{(t_{0})}
(t,\theta)+b_{k}^{*}(t_{0})\,G_{k}^{(t_{0})*}(t,\theta)
\right], \end{equation} where we have singled out the
homogeneous mode $k=0$ and used the symbol $*$ to
represent complex conjugation. The constants
$b_{k}(t_{0})$ are complex coefficients, $q_{0}$ and
$p_{0}$ are canonically conjugate variables and the
mode solutions $G_{k}^{(t_{0})}(t,\theta)$ are given
by
\begin{equation}
\label{to-modes}
G_{k}^{(t_{0})}(t,\theta)=\sqrt{\frac{\pi t}{4}}
\left[c^{*}\left(|k|t_{0}\right)
H_0(|k|t)-d^{*}\left(|k|t_{0}\right) H^*_0(|k|t)
\right]e^{*}_k,
\end{equation}
where
\begin{equation}
\label{candd} d(x)=\sqrt{\frac{\pi
x}{8}}\left[\left(1+\frac{i}{2x}
\right)H^{*}_{0}(x)-iH^{*}_{1}(x)\right],\quad c(x)=
\sqrt{\frac{\pi x}{2}}H_{0}(x)-d^{*}(x),
\end{equation}
and $H_{n}$ ($n=1,2$) is the $n$-th order Hankel
function of the second kind \cite{abra}. Note that the
mode solutions satisfy
\begin{equation}\label{Gvalues}
G_{k}^{(t_{0})}(t,\theta)\vert_{t=t_{0}}=\frac{e^{*}_k}
{\sqrt{2\vert k \vert}},\quad\quad
\partial_{t}G_{k}^{(t_{0})}(t,\theta)\vert_{t=t_{0}}=-
i\sqrt{\frac{\vert k \vert}{2}}e^{*}_{k}.\end{equation}

We will refer to the linear space of solutions
(\ref{solutions}), equipped with the symplectic
structure $\Omega\left(\xi_1,\xi_{2}\right)=\oint
\d\theta\left(\xi_{2}\partial_{t}\xi_{1}-
\xi_{1}\partial_{t}\xi_{2}\right)$, as the covariant
phase space $S$.

Alternatively, instead of $S$, we can consider the
canonical phase space. This is the linear  space
${\Gamma}$ coordinatized by the canonical pair which
is formed by the configuration $\varphi$ and the
momentum $P_{\varphi}$ of the field $\xi$ on a given
section of constant time. We take this time to be some
fixed reference time $t_0$. As we have seen above, the
section of constant time $t=t_0$ can be identified
with the compact space $S^1$. In the following, we
will refer to this section as the reference Cauchy
surface (RCS). Let us also point out that, via Eq.
(\ref{Gvalues}), one can understand the way in which
the solutions (\ref{solutions}) are expressed as being
specially adapted to the choice of RCS, or vice-versa
(given the RCS, such an adapted expression of the
solutions obviously simplifies the explicit form of
the map between $\Gamma$ and $S$). On the other hand,
the symplectic structure on $\Gamma$ is, of course,
\begin{equation}\sigma\left[(\varphi_1,P_{\varphi_{1}}),
(\varphi_2,P_{\varphi_{2}})\right]=\oint  \d\theta
\left(P_{\varphi_{1}}\varphi_2-P_{\varphi_{2}}\varphi_1
\right).\end{equation}

The evolution generated by the Hamiltonian
(\ref{hamil}) in the canonical phase space gives rise
to a 1-parameter family of symplectic linear
transformations $\tau_{(t_{f},t_{0})}:\Gamma \to
\Gamma$ (with $t_0$ fixed) as follows. An initial
state $(\varphi,P_{\varphi})$ at $t=t_{0}$ determines
a solution $\xi\in S$, which in turn determines a
canonical pair of fields $(\xi|_{t=t_f},
\partial_{t}\xi|_{t=t_f})$ for any value of $t_f$.
This pair is then naturally interpreted as new initial
data at $t=t_0$. More rigorously, we have a natural
1-parameter family of embeddings $E_{t}:S^1\to {\cal
M}^{(f)}$, together with a 1-parameter family of
isomorphisms $I_{E_{t}}$, mapping Cauchy data at
$E_{t}(S^1)$ into solutions. Then, the classical
evolution operator is
\begin{equation}\tau_{(t_f,t_{0})}
=\left(E_{t_0}^*\right)^{-1}E_{t_f}^*
I_{E_{t_f}}^{-1}I_{E_{t_0}},\end{equation} with
$E_{t}^*$ denoting the pull-back of the map $E_{t}$.
In this work we mostly ignore the distinction between
$S^1$ and our RCS, $E_{t_0}(S^1)$, so that $E_{t_0}$
is trivialized. In addition, note that the canonical
evolution maps provide the transformations
$I_{E_{t_0}}\tau_{(t_f,t_{0})}I_{E_{t_0}}^{-1} :S\to
S$ in the covariant phase space (this notion of time
evolution in the covariant description was employed in
Ref. \cite{TV}).

In order to present the evolution maps in explicit
form, it is convenient to use the Fourier components
of the field $\varphi$ and its momentum. We then
define
\begin{equation}\varphi_{k}:=\oint \d\theta\;
\varphi e_{k},\quad \quad P^{k}_{\varphi}:=\oint
\d\theta\; P_{\varphi}e^{*}_{k}.\end{equation} It is
clear from the form of the Hamiltonian (\ref{hamil})
that modes with different values of $|k|$ decouple.
Furthermore, from now on we will concentrate ourselves
on the infinite set of inhomogeneous modes $k\not =0$,
since no relevant aspect of our discussion depends on
the single zero mode (being single and decoupled, the
quantum treatment of this mode can be made
independently by standard methods, and included in the
final description by means of a tensor product).

Employing Eq. (\ref{Gvalues}), one can check that the
Fourier coefficients $\varphi_{k}$ and
$P_{\varphi}^{k}$  are related  to those appearing in
expression (\ref{solutions}) by
\begin{equation}
\label{ann-cre-like}
b_{k}(t_{0})=\frac{1}{\sqrt{2\vert k \vert}}
\left(\vert k \vert \varphi_{k} + iP^{-k}_{\varphi}
\right),\qquad b^{*}_{-k}(t_{0})=
\frac{1}{\sqrt{2\vert k \vert}}\left(\vert k
\vert \varphi_{k} - iP^{-k}_{\varphi}\right).
\end{equation}
We will adopt this convenient set of (complex)
variables as alternative coordinates in $\Gamma$. In
the following, to simplify the notation, we will let
$b_{k}$ and $b^{*}_{-k}$ denote the variables
$b_{k}(t_0)$ and $b^{*}_{-k}(t_0)$, respectively, and
collect them in the set of pairs
$\{(b_{k},b^{*}_{-k})\}$ with $k\in
\mathbb{Z}-\{0\}$.\footnote{Note that the pairs with
$k>0$ and $k<0$ are related by complex conjugation.}
It is then straightforward to check that each of the
considered pairs of variables decouples in the
evolution, so that the evolution transformations are
$2\times 2$ block-diagonal in these coordinates.

In more detail, time evolution $\tau_{(t_{f},t_{0})}$
maps $(b_{k}, b^{*}_{-k})$ to a new pair
$\left(b_{k}(t_f), b^{*}_{-k}(t_f)\right)$ [seen as
new data at $t=t_0$, related to the new configuration
and momentum of the field as in Eq.
(\ref{ann-cre-like})], such that
\begin{eqnarray}
\label{bs-evolution}
b_{k}(t_{f})&=&\alpha_{k}(t_{f},t_{0})b_{k}
+\beta_{k}(t_{f},t_{0})b^{*}_{-k},\nonumber \\
b^*_{-k}(t_{f})&=&\beta^*_{k}(t_{f},t_{0})b_{k}
+\alpha^*_{k}(t_{f},t_{0})b^{*}_{-k},
\end{eqnarray}
where
\begin{eqnarray}
\label{bogo-coeff} \alpha_{k}(t_{f},t_{0})&=& c(\vert
k\vert t_{f})c^{*} (\vert k\vert t_{0})-d(\vert k\vert
t_{f})d^{*} (\vert k\vert t_{0}),\nonumber \\
\beta_{k}(t_{f},t_{0}) &=&d(\vert k\vert t_{f})c(\vert
k\vert t_{0})- c(\vert k\vert t_{f})d(\vert k\vert
t_{0}).
\end{eqnarray}
Note that the functions $c$ and $d$, given in Eqs.
(\ref{candd}), satisfy $\vert c\vert^{2}-\vert
d\vert^{2}=1$, and we thus have that $\vert
\alpha_{k}(t_{f},t_{0})\vert^{2}-\vert
\beta_{k}(t_{f},t_{0})\vert^{2}=1$ for all $t_{f}>0$
(and any $t_{0}>0$). In addition,
\begin{eqnarray}
\alpha_{-k}(t_{f},t_{0})&=&
\alpha_{k}(t_{f},t_{0}),\quad \quad
\beta_{-k}(t_{f},t_{0})=
\beta_{k}(t_{f},t_{0}),\nonumber\\
\label{relalbe} \alpha_{k}(t_{0},t_{f})&=&
\alpha^{*}_{k}(t_{f},t_{0}),\quad \quad
\beta_{k}(t_{0},t_{f})=-\beta_{k}(t_{f},t_{0}).
\end{eqnarray}

\subsection{Fock quantization}
\label{ccmv-quant-2}

Let us summarize now the Fock quantization of the
model, i.e. the Fock quantization of the sector of
nonzero modes of the associated scalar field system,
as performed in Refs. \cite{ccm1,ccm2}. We will call
this sector of nonzero modes the inhomogeneous sector.

By construction, the set of mode solutions
$\{G_{k}^{(t_{0})}(t,\theta),
G_{k}^{(t_{0})*}(t,\theta)\}$ in Eq. (\ref{solutions})
(with $k\in\mathbb{Z} -\{0\}$) is complete in the
inhomogeneous sector of the space of solutions $S$,
and ``orthonormal'' in the product
$(G_{k}^{(t_{0})},G_{m}^{(t_{0})})=
-i\Omega(G_{k}^{(t_{0})*},G_{m}^{(t_{0})})$, in the
sense that \begin{equation}
(G_{k}^{(t_{0})},G_{m}^{(t_{0})})=\delta_{km},\quad
\quad(G_{k}^{(t_{0})*},G_{m}^{(t_{0})*})=-\delta_{km},
\quad\quad(G_{k}^{(t_{0})},G_{m}^{(t_{0})*})=0.
\end{equation}
Associated to the field decomposition
(\ref{solutions}), there is a natural
$\Omega$-compatible complex structure $J_0$:
\begin{equation}
\label{fock-cs}
J_{0}\left[G_{k}^{(t_{0})}(t,\theta)\right]=
iG_{k}^{(t_{0})}(t,\theta),\qquad J_{0}
\left[G_{k}^{(t_{0})*}(t,\theta)\right]=
-iG_{k}^{(t_{0})*}(t,\theta).
\end{equation}
This complex structure defines (and is defined by) the
annihilation and creation-like variables
$b_k(t_0)=\Omega(J_0 G_{k}^{(t_0)*},\xi)$ and
$b_k^{*}(t_0)=\Omega(J_0 G_{k}^{(t_0)},\xi)$. We
notice that $J_0$ is invariant under the group of
$S^1$ translations $T_{\omega}:\theta \mapsto
\theta+\omega$ generated by the global constraint
(\ref{homo-const}).

Starting with $(S,J_0)$,  we can construct the
so-called ``one particle'' Hilbert space
${\cal{H}}_0$. It is the Cauchy completion of the
space of ``positive'' frequency solutions
\begin{equation}
S^{+}:=\left\{\xi^{+}=\frac{1}{2}(\xi-
iJ_{0}\xi)\right\}\end{equation} with respect to the
norm $\vert \vert \xi^{+}\vert \vert = \sqrt{\langle
\xi^{+},\xi^{+}\rangle}$. Here,
$\langle\cdot,\cdot\rangle$ denotes the inner product
$\langle \xi_{1}^{+},\xi_{2}^{+}\rangle :=
-i\Omega(\xi^{-}_1,\xi^{+}_2)$ with
$\xi^{-}=(\xi+iJ_{0}\xi)/2\in \bar{\cal{H}}_0$ (the
complex conjugate space of ${\cal{H}}_0$). The
kinematical Hilbert space of the quantum theory is
then the symmetric Fock space
\begin{equation}
\label{fock-space}
{\cal{F}}({\cal{H}}_{0})=\bigoplus_{n=0}^{\infty}
\left(\bigotimes\,_{(s)}^{n} {\cal{H}}_{0}\right),
\end{equation}
where $\otimes\,_{(s)}^{n} {\cal{H}}_{0}$ is the
Hilbert space of all $n$-th rank symmetric tensors
over ${\cal{H}}_0$. Following this prescription, the
formal field operator $\hat{\xi}$ yields
\begin{equation}
\label{field-op} \hat{\xi}(t;\theta)=\sum_{k\in
\mathbb{Z}-
\{0\}}\left[G_{k}^{(t_{0})}(t,\theta)\hat{b}_{k}
+G_{k}^{(t_{0})*}(t,\theta)\hat{b}_{k}^{\dagger}
\right].
\end{equation}
Here, $\hat{b}_{k}$ and $\hat{b}_{k}^{\dagger}$ are,
respectively, the annihilation and creation operators
corresponding to the ``positive'' and ``negative''
frequency decomposition defined by $J_0$, and
represent the classical variables $b_{k}$ and
$b_{k}^{*}$.

A crucial aspect of this  quantization is that the
dynamics is unitarily implementable, i.e. for each
symplectic transformation in the 1-parameter family
$\tau_{(t_f,t_{0})}$ (\ref{bs-evolution}) defined by
time evolution $\forall t_f>0$, there exists a unitary
quantum evolution operator $\hat{U}(t_f,t_0)$ such
that
\begin{eqnarray}
\hat{b}_{k}(t_f)&=&\alpha_{k}(t_f,t_{0})
\hat{b}_{k}+\beta_{k}(t_f,t_{0})
\hat{b}^{\dagger}_{-k}=
\hat{U}^{-1}(t_f,t_0)\hat{b}_{k} \hat{U}(t_f,t_0),
\nonumber
\\ \hat{b}^{\dagger}_{-k}(t_f)&=&
\beta^*_{k}(t_f,t_{0})\hat{b}_{k}+
\alpha^*_{k}(t_f,t_{0})\hat{b}^{\dagger}_{-k}=
\hat{U}^{-1}(t_f,t_0)\hat{b}^{\dagger}_{-k}
\hat{U}(t_f,t_0). \label{q-evol}
\end{eqnarray}
As shown in Refs. \cite{ccm1,ccm2}, this follows from
the fact that the sequences $\{\beta_{k}(t_f,t_0)\}$
are square summable.\footnote{Let us recall that a
symplectic transformation is unitarily implementable
with respect to a FR if, and only if, its antilinear
part is Hilbert-Schmidt on the ``one particle''
Hilbert space \cite{shale}. In the present case this
condition reduces to $\sum_{k}\vert
\beta_{k}(t_f,t_0)\vert^{2}<\infty$.}

In addition, since $J_0$ is invariant under the group
of translations $T_{\omega}$, we have an invariant
unitary implementation of the gauge group on the
(kinematical) Fock space ${\cal{F}}({\cal{H}}_{0})$.

The physical Hilbert space ${\cal{F}}_{{\rm{phys}}}$
consists of all states in ${\cal{F}}({\cal{H}}_{0})$
that belong to the kernel of the quantum constraint
\begin{equation}
\label{const-fock}
{\hat{\cal{C}}}_{0}=\sum_{k=1}^{\infty}k
\left(\hat{b}^{\dagger}_{k}\hat{b}_{k}
-\hat{b}^{\dagger}_{-k}\hat{b}_{-k}\right).
\end{equation}
Starting with the basis of ``$n$-particle'' states
determined by the annihilation and creation operators
$\{(\hat{b}_{k},\hat{b}^{\dagger}_{k})\}$, one can
then construct physical states by restricting the
elements of that basis to the subset of states which
are physical, namely, the ``$n$-particle'' states with
zero field momentum
$\sum_{k=1}^{\infty}k(N_k-N_{-k})=0$, where $N_k$ is
the corresponding eigenvalue of the {\sl partial}
$k$-th number operator
$\hat{N}_k:=\hat{b}^{\dagger}_{k}\hat{b}_{k}$.
Furthermore, it is straightforward to check that
${\hat{\cal{C}}}_{0}$ is invariant under the time
evolution (\ref{q-evol}). This invariance ensures that
the dynamics is unitarily implementable not just on
${\cal{F}}({\cal{H}}_{0})$, but also on the space of
physical states ${\cal{F}}_{{\rm{phys}}}$.

Let us conclude with a comment regarding an apparent
ambiguity. Our fixed reference time $t_0$ certainly
plays a role in the definition of $J_0$, and it is
clear that, by changing $t_0$ and keeping the
definition (\ref{fock-cs}), one obtains new complex
structures with the same properties of
$S^1$-invariance and unitary dynamics, since the
results of Refs. \cite{ccm1,ccm2} do not depend on the
value of $t_0$. However, since these different complex
structures are, by construction, related by evolution
transformations, they give rise to unitarily
equivalent quantizations, precisely because the
evolution is unitary \cite{ccm2} (see also Subsec.
\ref{evolution}). Moreover, as we mentioned in the
introduction, much stronger results have indeed been
proven regarding the uniqueness of the quantization
\cite{ccmv,cmv}.

\section{The Schr\"{o}dinger representation}
\label{sec:3}

We will now  obtain the Schr\"{o}dinger functional
description of the quantum representation of the
canonical commutation relations (CCRs) provided by the
quantum fields of the system at a given fixed time.
Let us stress again that, just because of the unitary
implementation of the field dynamics, the choice of
this fixed time is irrelevant, in the sense that
different choices lead to unitarily equivalent
representations of the CCRs. So, for convenience, we
will take this fixed time to be our reference time
$t_0$.

The SR that we are going to construct is that defined
by the specific complex structure that is induced from
$J_0$ on the canonical phase space $\Gamma$ by means
of the isomorphism $I_{E_{t_0}}$. Taking into account
that the complex structure $J_0$ effectively declares
that the classical variables $\{b_k\}$ and $\{b_k^*\}$
are to be quantized as the respective annihilation and
creation operators of the representation, and
recalling Eq. (\ref{ann-cre-like}), which gives the
relation between these variables and the field modes,
it should not come as a surprise that the
representation of the CCRs which we will obtain is
essentially that associated with the free massless
field in $S^1$. We will nevertheless present this
construction in some detail, both for completeness and
to clarify the relation that, for the quantization of
the Gowdy model, exists between the covariant approach
adopted in Refs. \cite{ccm1,ccm2} and its canonical
version.

\subsection{General framework}

Let us start by considering the canonical phase space
$\Gamma$ (more precisely, its inhomogeneous sector).
The set of elementary observables ${\cal{O}}$ is taken
to be the vector space of linear functionals
\begin{equation}
\label{obs}
L_{\lambda}(Y):=\sigma\left(\lambda,Y\right)= \oint
\d\theta\left(f\,\varphi + g\,P_{\varphi}\right)
\end{equation}
and the unit functional ${\bf 1}$, namely ${\cal{O}}=
{\rm{Span}}\{{\bf 1},L_{\lambda}\}$. Here, $Y$ is a
vector in $\Gamma$ of the form $(\varphi,P_{\varphi})$
and $\lambda$ denotes a pair of smooth test functions
$(-g,f)$ which have both a vanishing integral on
$S^1$. The set $\cal{O}$ is closed under Poisson
brackets,
$\{L_{\lambda}(Y),L_{\nu}(Y)\}=L_{\nu}(\lambda)$, and
is complete, in the sense that its elements separate
points in (the inhomogeneous sector of) $\Gamma$.

The configuration and momentum observables are
particular cases of functionals $L_{\lambda}$. Whereas
$L_{\lambda}\vert_{\lambda=(0,f)}$ defines the
configuration observable{\footnote{The Fourier
components of $f$ and $g$ in $\lambda$ are,
respectively, $f^{k}=\oint \d\theta\, f e^{*}_{k}$ and
$g_{k}=\oint \d\theta\; g e_{k}$.}}
\begin{equation}
\label{conf:obs} \bar{\varphi}(f):= \oint \d\theta\,
f\varphi=\sum_{k\in \mathbb{Z}-\{0\}}f^{k}\varphi_{k},
\end{equation}
the momentum observable is defined by considering the
label $\lambda=(-g,0)$,
\begin{equation}
\label{mom:obs} \bar{P}_{\varphi}(g):= \oint \d\theta\,
g\,P_{\varphi}=\sum_{k\in \mathbb{Z}-\{0\}}g_{k}
P_{\varphi}^{k}.
\end{equation}
{}From the Poisson brackets between the configuration
and momentum observables (and setting $\hbar=1$), one
obtains for their respective quantum operators
$\hat{\bar{\varphi}}[f]$ and
$\hat{\bar{P}}_{\varphi}[g]$ the CCRs:
\begin{equation}\left[\hat{\bar{\varphi}}[f],
\hat{\bar{P}}_{\varphi}[g]\right] = i\hat{\bf 1}
\sum_{k\in \mathbb{Z}-\{0\}}f^{k}g_{k}.
\label{z4}\end{equation}

At this point of the discussion and in order to make
the analysis self-contained, it is convenient to
succinctly review how a Schr\"{o}dinger functional
representation of the CCRs is determined by a complex
structure on the canonical phase space. We will start
by describing the most general form of a complex
structure on $\Gamma$. This discussion can then be
easily applied to the general setting of a scalar
field in a globally hyperbolic spacetime (see Refs.
\cite{ccq-ap,ash-mag}) and, in particular, to the case
of the Gowdy model.

A ($\sigma$-compatible) complex structure $j$ on
$\Gamma$ has the generic form
\begin{equation}
\label{general-cs}
j(\varphi,P_{\varphi})=\left(A\varphi+
B P_{\varphi},CP_{\varphi}+D\varphi\right),
\end{equation}
where $A$, $B$, $C$ and $D$ are linear operators that
satisfy
\begin{eqnarray}
\label{cs:condition} A^{2}+BD=-{\bf{1}},&&
AB+BC=0, \nonumber \\ C^{2}+DB=-{\bf{1}},&&
DA+CD=0
\end{eqnarray}
(so that $j^2=-{\bf 1}$), and
\begin{eqnarray}
\label{cs:compatible}
(f,Bf')=(Bf,f'),\quad \quad (g,Dg')=(Dg,g'),\nonumber \\
(f,Ag)=-(Cf,g), \quad \quad(f,Bf)<0,\quad \quad
(g,Dg)>0
\end{eqnarray}
for all smooth test functions
 $g$, $g'$, $ f$ and $f'$ (so
that $j$ is $\sigma$-compatible). Here, we have
introduced the notation $(f,g):=\oint \d\theta\,f g$.
Notice that $C$ and $D$ can be obtained from $A$ and
$B$: indeed, from the two first relations in Eq.
(\ref{cs:condition}) one gets $C=-B^{-1}\,A\,B$ and
$D=-B^{-1}({\bf 1}+A^2)$ (when $B^{-1}$ exists). Thus,
the set of all compatible complex structures on
$\Gamma$ can be parameterized by the operators $A$ and
$B$ (assuming $B$ is invertible); that is, this set
can be identified with $\{j_{(A,B)}\}$ where (in
matrix notation)
\begin{equation}
\label{gen-para-cs}
j_{(A,B)}=\left(\begin{array}{cc} A & B \\
-B^{-1}({\bf 1}+A^2) &  \quad -B^{-1}AB \end{array}
\right).
\end{equation}

Given a complex structure $j$ on the canonical phase
space $\Gamma$, a Schr\"{o}dinger, or
``configuration'' wave functional representation --
which we will call the $j$-SR -- is determined as
follows.\footnote{We are only presenting the outcome,
obtained under suitable regularity conditions. The
full process involves the construction of an inner
product from $j$ and $\sigma$, which is used to
determine a state of the Weyl algebra associated with
the CCRs. The GNS representation defined by this state
can be realized as an SR, since the restriction of the
state to the Weyl configuration observables defines a
measure.} The $j$-SR consists of a representation of
the basic operators of configuration and momentum on a
space of complex-valued functionals $\Psi$ on the
``quantum'' configuration space ${\bar{\cal{C}}}$
(generally an extension of the classical configuration
space). These functionals are square integrable with
respect to a Gaussian measure $\mu$ with covariance $-
B/2$.\footnote{We define the covariance of a Gaussian
measure as twice the positive bilinear form appearing
in the exponential of the Fourier transform of the
measure. We follow the standard practice of using the
term ``covariance'' to refer not only to this bilinear
form, but also to the operator which defines it with
respect to a fiducial integration in the space of test
functions, which in our case is given by $\d\theta$.}
On the Hilbert space defined in this way, the basic
operators of configuration and momentum are
\begin{eqnarray}
\left(\hat{\bar{\varphi}}[f]\Psi
\right)[\bar{\varphi}]&=& \bar{\varphi}(f)
\Psi[\bar{\varphi}],
\\ \label{gen-rep-mom}
\left(\hat{\bar{P_{\varphi}}}[g]\Psi\right)
[\bar{\varphi}]&=&-i\frac{\delta\Psi}
{\delta{\bar{\varphi}}}[g] -i\bar{\varphi}
\left(B^{-1}({\bf{1}}-iA)g\right)\Psi[\bar{\varphi}],
\end{eqnarray}
where $\bar{\varphi}\in \bar{\cal{C}}$.

It is worth noticing that, while the measure is
determined just by $B$, there is an extra freedom in
the momentum operator, given by the operator $A$ (see
Ref. \cite{ccq-cqg} for discussion). Finally, let us
also recall that two complex structures $j$ and $j'$
on $\Gamma$ lead to unitarily equivalent
representations of the CCRs if, and only if, $j-j'$
defines a Hilbert-Schmidt operator on the ``one
particle'' Hilbert space determined by $j$ (or
equivalently by $j'$).

\subsection{The canonical complex structure}

As we explained in Sec. \ref{ccmv-quant}, given our
RCS, which is determined by the chosen reference time
$t_0$, there is a preferred isomorphism between the
canonical phase space and the space of solutions to
the field equation (\ref{field-eq}). In order to
simplify the notation, we will denote this isomorphism
by $I_{E_0}$ instead of $I_{E_{t_0}}$. Then,
$I_{E_0}:\Gamma\to S$ is such that
\begin{equation}
\label{z1} S\ni\xi\mapsto
I_{E_0}^{-1}(\xi)=(\varphi,P_{\varphi})=
(\xi|_{t=t_0},\partial_t\xi|_{t=t_0}).
\end{equation}
Therefore, a complex structure $J$ on the covariant
phase space $S$ determines (and is determined by) a
corresponding complex structure $j=I_{E_0}^{-1} J
I_{E_0}$ on the canonical phase space. In particular,
the complex structure $J_0$ of Sec. \ref{ccmv-quant}
has the canonical counterpart $j_0=I_{E_0}^{-1} J_0
I_{E_0}:\Gamma\to\Gamma$. The SR we are looking for is
thus specified by $j_0$, following the prescription of
the previous subsection. We will now obtain the
explicit form of $j_0$.

Recalling the field decomposition (\ref{solutions})
(for the inhomogeneous sector) and employing Eq.
(\ref{Gvalues}), we get the explicit relation between
($\varphi$, $P_{\varphi}$) and the set of pairs of
variables $\{(b_k,b_{-k}^{*})\}$:
\begin{equation} \label{conf-to}
\varphi= \sum_{k\in \mathbb{Z}-\{0\}}\frac{1}
{\sqrt{2\vert k \vert}}\left[
b_{k}\,e^{*}_k+b_{k}^{*}\, e_k\right],\quad \quad
P_{\varphi}= -i\sum_{k\in
\mathbb{Z}-\{0\}}\sqrt{\frac{\vert
k\vert}{2}}\left[b_{k}\,e^{*}_k
-b_{k}^{*}\,e_k\right].
\end{equation}
For a given $(\varphi,P_{\varphi})\in\Gamma$ and the
corresponding solution
$\xi=I_{E_0}(\varphi,P_{\varphi})\in S$, we obtain the
new canonical fields $j_{0}(\varphi,P_{\varphi})=
I_{E_0}^{-1}J_0(\xi)\in\Gamma$, which we will call
$(\tilde{\varphi},\tilde{P}_{\varphi})$. Taking into
account that $J_{0}(\xi)=i\xi^{+}-i\xi^{-}$, with
$\xi^{+}$ ($\xi^{-}$) being the ``positive''
(``negative'') frequency part spanned by
$\{G_{k}^{(t_{0})}\}$ ($\{G_{k}^{(t_{0})*}\}$), with
$k\in \mathbb{Z}-\{0\}$, we get
\begin{equation}\tilde{\varphi}=
J_0({\xi})\vert_{t_{0}}=
i\xi^{+}\vert_{t_{0}}-i\xi^{-} \vert_{t_{0}},\quad
\tilde{P}_{\varphi}=\partial_{t}J_0({\xi})
\vert_{t_{0}}=i\partial_{t}\xi^{+}\vert_{t_{0}}
-i\partial_{t} \xi^{-}\vert_{t_{0}}.
\end{equation}
Hence, it is easy to check that
\begin{equation}
\tilde{\varphi}= \sum_{k\in
\mathbb{Z}-\{0\}}\frac{i}{\sqrt{2 \vert k \vert}}
\left[b_{k}\,e^{*}_k-b_{k}^{*}\, e_k\right], \quad
\quad \label{mom-j} \tilde{P}_{\varphi}=
\sum_{k\in\mathbb{Z}-\{0\}}\sqrt{\frac{\vert k
\vert}{2}} \left[b_{k}\,e^{*}_k+b_{k}^{*}
\,e_k\right].\end{equation} From Eqs. (\ref{conf-to})
and (\ref{mom-j}), one obtains that
$\tilde{\varphi}=-(-\Delta)^{-1/2}\,P_{\varphi}$ and
$\tilde{P}_{\varphi}=(-\Delta)^{1/2}\,\varphi$, where
$\Delta$ is the second order differential operator
$\d^{2}/\d\theta^{2}$. The explicit expression for the
canonical counterpart of $J_{0}$ is then
\begin{equation}
\label{canonical-cs} j_{0}=\left(  \begin{array}{cc} 0
& -(-\Delta)^{-1/2} \\ (-\Delta)^{1/2} & 0
\end{array} \right).
\end{equation}
A comparison with Eq. (\ref{gen-para-cs}) shows that,
in this case, $A=0$ and $B=-(-\Delta)^{-1/2}$.
Therefore, the momentum operators are completely
determined by the covariance $(-\Delta)^{-1/2}/2$ of
the Gaussian measure.

In terms of the Fourier coefficients $\{(\varphi_{k},
P_{\varphi}^{-k})\}$ with $k\in\mathbb{Z}-\{0\}$, the
complex structure (\ref{canonical-cs}) yields
\begin{equation}
\label{ccs-fourier} (j_{0})_k=\left(
\begin{array}{cc} 0 & -\frac{1}{|k|}  \\ |k| & 0
\end{array} \right).
\end{equation}
So, in this alternative description of $\Gamma$
provided by the Fourier components of $\varphi$ and
$P_{\varphi}$, the counterparts of $A$ and $B$ are
given by $A_{k}=0$ and $B_{k}=-\frac{1}{|k|}$,
respectively (recall that $k\neq 0$).

\subsection{The functional representation of the
Gowdy cosmologies}
\label{subsec:fr}

Let us now complete the construction of the
$j_{0}$-SR. We will call ${\cal{T}}$ our space of test
functions, i.e. the space of smooth real functions on
$S^{1}$ with vanishing integral. By standard arguments
in the theory of measures in infinite dimensional
spaces (see e.g. Ref. \cite{unp-ashtekar}), the space
${\cal{T}}$ can be equipped with a so-called nuclear
topology, and the covariance $(-\Delta)^{-1/2}/2$
defines a Gaussian measure $\mu$ on the topological
dual of $\cal{T}$, namely the real vector space
${\cal{T}}^{\star}$ of continuous linear functionals
on $\cal{T}$. This will be the quantum configuration
space $\bar{\cal{C}}$.

Designating a generic element of ${\cal{T}}^{\star}$
as $\bar \varphi$ and its action on elements of
$\cal{T}$ as $f\mapsto\bar \varphi(f)$, the measure
$\mu$ is defined by its Fourier transform
\begin{equation}
\label{z3} \int_{{\cal{T}}^{\star}} e^{i\bar
\varphi(f)} \d\mu=\exp\left[-\frac{1}{4}
(f,(-\Delta)^{-1/2}f)\right].
\end{equation}
The ``configuration'' wave functional representation
of $\hat{\bar{\varphi}}$ and $\hat{\bar{P}}_{\varphi}$
on ${\cal{H}}_{s} :=L^{2}({\cal{T}}^{\star},\d\mu)$ is
then
\begin{eqnarray}
\label{jo-confrep} \left(\hat{\bar{\varphi}}[f]
\Psi\right)[\bar{\varphi}]
&=&\bar{\varphi}(f)\Psi[\bar{\varphi}],\\
\left(\hat{\bar{P}}_{\varphi}[g]
\Psi\right)[\bar{\varphi}] \, &=& -i\frac{\delta
\Psi}{\delta \bar{\varphi}}[g]
+i\bar{\varphi}((-\Delta)^{1/2}g)\Psi[\bar{\varphi}]
\label{jo-momrep}.
\end{eqnarray}

An alternative description is obtained in Fourier
space as follows. By means of the Fourier
correspondence $f\mapsto \{f^{k}\}=\{\oint \d\theta f
e^{*}_k\}$, one can identify ${\cal{T}}$ with the
space of rapidly decreasing complex sequences
$\{f^k\}$ with $k\in \mathbb{Z}-\{0\}$, i.e. sequences
such that $k^{r}f^{k}$ goes to zero as $|k|\to
\infty$, for all $r>0$ (and which, moreover, satisfy
${f^k}^*=f^{-k}$, so that the corresponding functions
$f$ are real). Likewise, the dual space
${\cal{T}}^{\star}$ can be identified with a subspace
(of sequences of appropriate behavior) of the space of
all complex sequences $\{\varphi_k\}$ with $k\in
\mathbb{Z}-\{0\}$ and $\varphi_{-k}=\varphi_{k}^{*}$.
This correspondence is given by
${\cal{T}}^{\star}\ni\bar \varphi \leftrightarrow
\{\varphi_{k}\}:=\{\bar{\varphi}(e_{k})\}$, so that
\begin{equation}\bar{\varphi}(f)=\sum_{k\neq
0}f^{k}\varphi_{k}=\sum_{k>0}f^{k}\varphi_{k}+
\sum_{k>0}\left(f^{k}\varphi_{k}\right)^*.
\end{equation}
In order to present the measure without unnecessary
complications, we note that, since the sequences
$\{f^k\}\in {\cal{T}}$ and
$\{\varphi_k\}\in{\cal{T}}^{\star}$ are both
determined by their values for $k>0$, one can simply
work with sequences whose index is defined in
$\mathbb{N}$, rather than in $\mathbb{Z}-\{0\}$.
Actually, one can view $\mu$ as a measure on the space
of all complex sequences $\{\varphi_k\}$ with $k\in
\mathbb{N}$ that happens to be supported on the
subspace ${\cal{T}}^{\star}$.\footnote{On the other
hand, one can certainly find proper subsets of
${\cal{T}}^{\star}$ which support the measure. See
e.g. Ref. \cite{MTV} for reviews of results and for
techniques concerning support properties of field
measures.}

In this description, $\mu$ is a product measure on (a
subset of) the product space $\mathbb{C}^{\mathbb{N}}$
of complex sequences $\{\varphi_k\}$ with $k\in
\mathbb{N}$:
\begin{equation}
\label{measure-b} \d\mu = \prod_{k\in\mathbb{N}}
\frac{2\vert k \vert}{\pi}\exp\left(-2\vert k\vert\,
\vert \varphi_{k}\vert^{2}\right) \d\mu_k^0,
\end{equation}
where $\d\mu_k^0$ is the Lebesgue measure on the plane
coordinatized by $(\varphi_k,\varphi_k^*)$. It is
easily seen that this measure corresponds to that
appearing in Eq. (\ref{z3}).

Note that we are using here complex canonical
variables. This accounts for the factors 2 in Eq.
(\ref{measure-b}), which no longer appear when the
quantization is recasted in terms of real canonical
variables, namely the coefficients in the Fourier
decompositions of $\varphi$ and $P_{\varphi}$ in terms
of normalized sine and cosine functions.

It is worth pointing out that one can reinterpret the
measure $\mu$ described above as a measure on the
original space of sequences $\{\varphi_k\}$ with
integer index ($k\in\mathbb{Z}-\{0\}$) and such that
$\varphi_{-k}=\varphi_{k}^{*}$. Using the one-to-one
correspondence between these sequences and their
restrictions to $k\in \mathbb{N}$, one can define both
the measurable sets and the measure.

The operators which present the simplest expressions
correspond to the Fourier components of the field
operators, $\hat{\bar{\varphi}}[e_{k}]$ and
$\hat{\bar{P}}_{\varphi}[e^{*}_{k}]$, i.e. to the
quantization of the classical variables $\varphi_k$
and $P_{\varphi}^k$:
\begin{eqnarray}
\label{jo-fc-crep}
\hat{\varphi}_{k} \Psi &=& \varphi_{k}\Psi, \\
\hat{P}_{\varphi}^{k} \,\Psi &=& -i\frac{\partial
\Psi}{\partial \varphi_{k}}+i\vert k \vert
\varphi_{-k}\Psi, \label{jo-fc-mrep}
\end{eqnarray}
where $\Psi$ is a functional of the Fourier components
$\varphi_{k}$.

The CCRs (\ref{z4}) are clearly satisfied. Moreover,
the same happens with the reality conditions
$\hat{\varphi}_{k}^{\dagger}=\hat{\varphi}_{-k}$ and
$\hat{P}_{\varphi}^{k\dagger}=\hat{P}_{\varphi}^{-k}$
with respect to the
$L^2({\cal{T}}^{\star},\d\mu)$-inner product.
Equivalently, the operators $\hat{\bar{\varphi}}[f]$
and $\hat{\bar{P}}_{\varphi}[g]$ are symmetric,
leading to self-adjoint operators on an appropriate
domain of definition.

In addition, from Eq. (\ref{ann-cre-like}) the
variables ${b}_{k}$ and ${b}_{k}^{*}$ are quantized as
\begin{eqnarray}
\hat{b}_{k}&=&\frac{1}{\sqrt{2\vert k\vert}}
\frac{\partial}{\partial \varphi_{-k}},\nonumber \\
\hat{b}_{k}^{\dagger}&=&- \frac{1}{\sqrt{2\vert
k\vert}}\frac{\partial}{\partial
\varphi_{k}}+\sqrt{2\vert k\vert} \varphi_{-k}.
\label{ann-crea-op}
\end{eqnarray}
These are precisely the  annihilation and creation
operators of the $j_{0}$-SR. By construction, the
``zero particle'' state of the $j_{0}$-SR, which we
will call the vacuum, is the unit constant functional
$\Psi_{0}[\bar{\varphi}]=1$ (up to a constant phase).

As we have already mentioned, the invariance of $J_0$
-- and therefore of $j_0$ -- under the group of
$S^{1}$-translations $T_{\omega}:\theta \to
\theta+\omega$,  $\omega \in S^{1}$, provides us with
corresponding unitary operators $\hat{T}_{\omega}$
which leave the vacuum invariant, and whose explicit
action, in the Fourier description, is given by
\begin{equation}
\label{quant-gauge-impl}
\left(\hat{T}_{\omega}\Psi\right)[\varphi_k]=
\Psi[e^{-i\, \omega \,k}\varphi_{k}], \ \ \ \Psi \in
{\cal{H}}_{s}.
\end{equation}
The generator of the unitary group $\hat{T}_{\omega}$,
\begin{equation}
\label{const-in-fp}
\hat{{\cal{C}}}_0=\sum_{k=1}^{\infty}\vert k
\vert\left(\varphi_{-k}\frac{\partial}{\partial
\varphi_{-k}}-\varphi_{k}\frac{\partial}{\partial
\varphi_{k}}\right),
\end{equation}
is the quantum constraint operator in the functional
approach.

The space of physical states consists of all states in
${\cal{H}}_{s}$ which are invariant under the action
of $\hat{T}_{\omega}$ for every $\omega \in S^{1}$.
That is, physical states are invariant under the group
of phase transformations $\varphi_{k}\to e^{-i\,
\omega \,k}\varphi_{k}$ $\forall\omega \in S^{1}$.
This property allows a characterization of physical
states alternative to that presented at the end of
Sec. II. One can obtain the Hilbert space of physical
states ${\cal{H}}_{\rm{phys}}$ as the quotient of the
kinematical Hilbert space ${\cal{H}}_{s}$ by the
action of the considered gauge group. Since this group
is compact, the projection of any kinematical state
$\Psi$ onto the space of physical states can then be
easily determined by a group averaging procedure (see
e.g. Ref. \cite{group-av}):
\begin{equation}
\Psi_{\rm{phys}}[\varphi_k]=\oint \frac{\d\omega}{2\pi}
\left(\hat{T}_{\omega}\Psi\right)
[\varphi_k].\end{equation} It is important to
emphasize that, because the gauge group is unitary and
compact, the physical state $\Psi_{\rm{phys}}$ has a
finite norm for any $\Psi\in {\cal{H}}_{s}$.
Therefore, the space of physical states is just a
Hilbert subspace of the kinematical Hilbert space.

In summary, the $j_{0}$-SR consists of a (kinematical)
Hilbert space ${\cal{H}}_{s}$ defined by a Gaussian
measure of covariance $(-\Delta)^{-1/2}/2$, on which
the CCRs are implemented by the operators
$\hat{\bar{\varphi}}$ and $\hat{\bar{P}}_{\varphi}$
(\ref{jo-confrep})-(\ref{jo-momrep}) [or equivalently,
by $\hat{\varphi}_{k}$ and $\hat{P}_{\varphi}^{k}$
(\ref{jo-fc-crep})-(\ref{jo-fc-mrep})].  The physical
Hilbert space consists of the invariant subspace under
$S^1$-translations. It follows from the results of
Refs. \cite{ccm1,ccm2,ccmv,cmv} that the $j_0$-SR is
the (essentially) unique $S^1$-invariant
``configuration'' wave functional representation with
a unitary dynamics. Finally, it is worth emphasizing
that the SR here presented is not equivalent to the
Schr\"odinger representations (SRs) constructed in
Ref. \cite{torre-sr}, where the considered basic field
was $\bar{\xi}=\xi/\sqrt{t}$ instead of $\xi$
\cite{cmv,ccm1,ccm2}.

\section{Time evolution}
\label{sec:5}

In this section we will address the issue of how time
evolution is implemented in our model in the context
of the functional representation.

\subsection{Creation and annihilation operators and the vacuum}

In the $J_0$-Fock quantization, classical dynamics is
implemented in the Heisenberg picture by a unitary
operator $\hat{U}(t_f,t_0)$ relating annihilation and
creation operators at different times as in Eq.
(\ref{q-evol}). Recalling that
$\hat{U}^{-1}(t_f,t_0)=\hat{U}(t_0,t_f)$ and the last
two relations in Eq. (\ref{relalbe}), we can now
introduce the annihilation and creation operators
corresponding to evolution ``backwards in time'',
\begin{eqnarray}
\hat{\bar{b}}_{k}(t_f)&=&\hat{U}(t_f,t_0)\hat{b}_{k}
\hat{U}^{-1}(t_f,t_0)=\alpha_{k}^{*}(t_f,t_{0})
\hat{b}_{k}-\beta_{k}(t_f,t_{0})\hat{b}^{\dagger}_{-k},
\nonumber
\\ \hat{\bar{b}}^{\dagger}_{k}(t_f)&=&\hat{U}(t_f,t_0)
\hat{\bar{b}}^{\dagger}_{k}\hat{U}^{-1}(t_f,t_0)=
\alpha_{k}(t_f,t_{0})\hat{b}^{\dagger}_{k}
-\beta^{*}_{k}(t_f,t_{0})\hat{b}_{-k}.
\label{q-evolback}
\end{eqnarray}
Obviously, $\hat{\bar{b}}_k(t_0)$ and
$\hat{\bar{b}}_k^{\dagger}(t_0)$ coincide with
$\hat{b}_k$ and $\hat{b}_k^{\dagger}$, respectively.

Because of the mixing of annihilation and creation
operators, the Heisenberg vacuum state $\vert
0\rangle_{\rm{H}}$ which is annihilated by all the
operators $\hat{b}_k$ fails to be in the kernel of all
the time-evolved operators $\hat{\bar{b}}_{k}(t_f)$
for any $t_{f}\neq t_0$. Instead, these operators
annihilate the state
\begin{equation}
\label{vac-rel} \vert
0,t_f\rangle:=\hat{U}(t_f,t_0)\vert 0\rangle_{{\rm{H}}},
\end{equation}
which is just the time-evolved vacuum, i.e. the
counterpart of the state $\vert 0\rangle_{{\rm{H}}}$
in the Schr\"odinger picture. Of course, $\vert
0,t_0\rangle=\vert 0\rangle_{{\rm{H}}}$.

We will refer to $\vert 0,t_f\rangle$ and to states of
the form
\begin{equation}
\label{n-tf-states} \vert n,t_f\rangle =
\hat{\bar{b}}^{\dagger}_{k_1}(t_f)
\hat{\bar{b}}^{\dagger}_{k_2}(t_f)\dots
\hat{\bar{b}}^{\dagger}_{k_n} (t_f)\vert 0,t_f\rangle
\end{equation}
as the $t_f$-vacuum and the $t_f$ ``$n$-particle''
states, respectively. From Eqs. (\ref{q-evolback}) and
(\ref{n-tf-states}) one concludes that the $t_f$
``$n$-particle'' states are related with the
Heisenberg ``$n$-particle'' states $\vert
n\rangle_{\rm{H}}:=\vert n,t_0\rangle$ as follows
\begin{equation} \vert n,t_f\rangle = \hat{U}
\hat{b}^{\dagger}_{k_1} \hat{b}^{\dagger}_{k_2}
\dots\hat{b}^{\dagger}_{k_n}\hat{U}^{-1}\vert
0,t_f\rangle= \hat{U}\hat{b}^{\dagger}_{k_1}
\hat{b}^{\dagger}_{k_2}\dots \hat{b}^{\dagger}_{k_n}
\vert 0\rangle_{\rm{H}}=\hat{U}\vert n
\rangle_{\rm{H}},
\end{equation}
where we have used $\hat{U}$ as an abbreviation for
$\hat{U}(t_f,t_0)$. The $t_f$ ``$n$-particle'' states
are thus the result of evolving the states $\vert n
\rangle_{\rm{H}}$ from $t_0$ to $t_f$. Therefore, in
order to specify the evolution to time $t_f$ of all
Heisenberg ``$n$-particle'' states --and hence
determine the time evolution operator--, we only need
to supply the operators (\ref{q-evolback}). In this
respect, we note that an equivalent condition for
unitarity of the evolution to time $t_f$ is the
existence of a vector which is annihilated by all the
operators $\hat{\bar{b}}_{k}(t_f)$. If this vector
exists, then it is unique (up to a constant phase), so
that the considered annihilation operators contain
indeed all the necessary information to fix the
evolved vacuum (\ref{vac-rel}).

Turning back to the functional description, let us now
write the operators $\hat{\bar{b}}_{k}(t_f)$ and
determine the explicit form of the state
$|0,t_f\rangle$ in the $j_0$-SR. From Eqs.
(\ref{ann-crea-op}) and (\ref{q-evolback}) one
obtains\footnote{One may also obtain the operators
$\hat{\bar{b}}_{k}^{\dagger}(t_f)$ in the same way.}
\begin{equation}
\hat{\bar{b}}_{k}(t_f)=\frac{\alpha_k^*+\beta_k}
{\sqrt{2\vert k\vert}}\frac{\partial}{\partial
\varphi_{-k}}-\sqrt{2\vert k\vert} \beta_k\varphi_{k}.
\end{equation}
Here, $\alpha_k$ and $\beta_k$ denote
$\alpha_k(t_f,t_0)$ and $\beta_k(t_f,t_0)$,
respectively, a simplified notation that we will use
in the following. It is straightforward to see that,
formally, the solution of the set of conditions
$\hat{\bar{b}}_{k}(t_f)\Psi=0$ ($\forall k\in
\mathbb{Z}-\{0\}$) is given by
\begin{equation}
\label{jf-vs} \Psi_0^{(t_f)}:=\prod_{k\in\mathbb{N}}
\frac{1}{ \vert \alpha_k^{*}+\beta_k \vert}\exp\left(2
\vert k \vert \frac{\beta_{k}}{\alpha_{k}^{*}
+\beta_{k}}\vert \varphi_{k}\vert^{2}\right),
\end{equation}
where we have already normalized each of the factors
in the infinite product. Actually, owing to the
summability of the sequences $\{|\beta_k|^2\}$ (i.e.
thanks to unitarity), one can check that the
normalized sequence formed by the finite number of
factors $1\leq k \leq K$ with $K\in \mathbb{N}$ is a
Cauchy sequence in the $L^2({\cal
T}^{\star},\d\mu)$-norm. Hence, the $t_f$-vacuum
$|0,t_f\rangle$ in the $j_0$-SR is (up to a constant
phase) the state $\Psi_0^{(t_f)}$, rigorously defined
as the $L^2$-limit of the sequence of products with a
finite number of factors.

\subsection{Complex structures induced by time
evolution}
\label{evolution}

Regardless of its unitary implementability in the
quantum theory, the classical evolution, being defined
by a family of symplectic transformations, generates a
family of representations of the CCRs starting from a
given one. In the present case, this family of
representations is associated with the family of
complex structures
\begin{equation}
\label{z5} j_{t_f}:=\tau_{(t_f,t_0)}j_0
\tau_{(t_f,t_0)}^{-1}, \quad\quad
j_{t_f}:\Gamma\to\Gamma,
\end{equation}
obtained by evolving the complex structure $j_0$.
Here, $\tau_{(t_f,t_0)}$ is the classical evolution
operator for an arbitrary time $t_f>0$. Clearly, the
condition of unitary implementability of time
evolution in the $j_0$-representation translates into
the condition of unitary equivalence between that
representation and the representations defined by the
complex structures $j_{t_f}$, $\forall t_f
>0$. Thus, one can address the question of time
evolution by considering the representations
constructed from the 1-parameter family of complex
structures $j_{t_f}$. The relationship between the
members of this  family of representations provides us
with an alternative, equivalent description of the
time evolution. In the present case, given the unitary
implementability of the evolution, established in
Refs. \cite{ccm1,ccm2}, we obtain a family of
unitarily equivalent representations. In particular,
the family of SRs defined by the complex structures
$j_{t_f}$, which we will refer to as the family of
$j_{t_f}$-SRs, is associated with a family of mutually
absolutely continuous Gaussian measures.

Before determining explicitly the complex structures
$j_{t_f}$ and the corresponding $j_{t_f}$-SRs, we will
give an equivalent characterization of them which is
related to the discussion in the previous subsection.
Let us consider the set of (pairs of) coefficients
$\{(\bar{b}_{k}(t_{f}),\bar{b}^{*}_{k}(t_{f}))\}$
which is obtained from $\{(b_{k},b^{*}_{k})\}$ by
applying $\tau_{(t_f,t_0)}^{-1}$ [i.e. the relation
between the two sets is the direct classical
counterpart of Eq. (\ref{q-evolback})]. It is clear
that, when expressed in terms of the pairs
$\{(\bar{b}_{k}(t_{f}),\bar{b}^{*}_{k}(t_{f}))\}$, the
complex structure $j_{t_f}$ adopts the same form as
$j_0$ in terms of the pairs $\{(b_{k},b^{*}_{k})\}$
[namely, it is given by a block-diagonal matrix with
the $2\times 2$ blocks $(j_{t_f})_k={\rm
diag}(i,-i)$]. Therefore, the $j_{t_f}$ representation
is such that the classical variables which are
quantized as the creation and annihilation operators
are $\{\bar{b}^{*}_{k}(t_{f})\}$ and
$\{\bar{b}_{k}(t_{f})\}$, respectively, rather than
$\{b^{*}_{k}\}$ and $\{b_{k}\}$.\footnote{Let us point
out that a different but equivalent way to recast time
evolution is with the family of representations
arising from the set of complex structures
$\{\tilde{j}_{t_f}={\tau}^{-1}j_0{\tau}\}$. In that
case, the annihilation and creation-like variables
defined by $\tilde{j}_{t_f}$ are $\{b_{k}(t_f)\}$ and
$\{b^{*}_{k}(t_f)\}$, introduced in Eq.
(\ref{bs-evolution}).}

Returning to the covariant description for a moment,
the family $\{j_{t_f}\}$ determines a family of
complex structures on the covariant phase space via
the isomorphism $I_{E_0}$ (\ref{z1}). These are given
by $J_{t_f}=\bar{\tau}_{(t_f,t_0)}J_0
\bar{\tau}_{(t_f,t_0)}^{-1}=I_{E_0}j_{t_f}
I_{E_0}^{-1}$, where $\bar{\tau}_{(t_f,t_0)}=I_{E_0}
\tau_{(t_f,t_0)}I_{E_0}^{-1}$ is the classical
evolution map in covariant phase space. Just as $J_0$
is associated with the field decomposition
(\ref{solutions}), $J_{t_f}$ can be understood as
being associated with the decomposition
\begin{equation}\label{fd-tf}\xi(t,\theta)=
\sum_{k\neq 0}\left[
\bar{b}_{k}(t_{f})\,G_{k}^{(t_{f})}(t,\theta)+
\bar{b}_{k}^{*}(t_{f})\,
G_{k}^{(t_{f})*}(t,\theta)\right],\end{equation} where
$G_{k}^{(t_{f})}=
\bar{\tau}_{(t_f,t_0)}G_{k}^{(t_{0})}$ are the
time-evolved modes. One can thus see that, as
commented above, changing the time used to define our
fiducial complex structure on the covariant phase
space corresponds in fact to evolution.

\subsection{The family of unitarily equivalent
functional representations}

Explicit expressions for the complex structures
$j_{t_f}$ (\ref{z5}) are obtained quite
straightforwardly. Taking into account expression
(\ref{ccs-fourier}) for $j_0$, relations
(\ref{ann-cre-like}) and the evolution
(\ref{bs-evolution}), one concludes that $j_{t_f}$,
given in terms of the Fourier coefficients
$\{(\varphi_{k}, P_{\varphi}^{-k})\}$, is defined by
the following $2\times 2$ matrices:
\begin{equation}
\label{ind-ccs-fourier}
(j_{t_f})_k=\left(\begin{array}{cc}
2{\rm{Im}}(\alpha_{k}\beta_{k}) &
-\frac{|\alpha_{k}^{*}+\beta_{k}|^{2}}{|k|} \\
|k||\alpha_{k}^{*}-\beta_{k}|^{2} &
\quad-2{\rm{Im}}(\alpha_{k}\beta_{k})
\end{array} \right).
\end{equation}

One can now easily determine the corresponding family
of $j_{t_f}$-SRs. Comparing with the case
(\ref{ccs-fourier}) for $j_0$, and referring to the
general form (\ref{gen-para-cs}), we find a change in
the terms $B_k$, which now become
$B_k=-|\alpha_{k}^{*}+\beta_{k}|^{2}/|k|$ and
correspond to a new Gaussian measure. In addition, we
note the appearance of the term
$A_{k}=2{\rm{Im}}(\alpha_{k}\beta_{k})$ (owing to the
mixing between ``positive'' and ``negative'' frequency
parts during evolution). The respective contribution
${\bf 1}-iA_{k}$ in the general expression for the
momentum operators (\ref{gen-rep-mom}) can be written
in this case as
$(\alpha_{k}+\beta_{k}^{*})(\alpha_{k}^{*}-\beta_{k})$.
Thus,
\begin{equation}
B_{k}^{-1}({\bf
1}-iA_{k})=-|k|\frac{\alpha_{k}^{*}-\beta_{k}}
{\alpha_{k}^{*}+\beta_{k}}.\end{equation} Adopting the
same Fourier space description as in Subsec.
\ref{subsec:fr}, the $j_{t_f}$-SR is then realized in
the Hilbert space
$L^{2}({\cal{T}}^{\star},\d\mu_{t_f})$ defined by the
Gaussian product measure
\begin{equation}
\label{measure-b-tf} \d\mu_{t_f} =
\prod_{k\in\mathbb{N}}\frac{2|k|}{\pi
|\alpha_{k}^{*}+\beta_{k}|^{2}}
\exp\left(-\frac{2|k|}{|\alpha_{k}^{*}+\beta_{k}|^{2}}
\vert\varphi_{k}\vert^{2}\right) \d\mu_k^0,
\end{equation}
where $\d\mu_k^0$ is again the Lebesgue measure in
$\mathbb{C}$.

The (Fourier components of the) basic field operators
are now represented by
\begin{eqnarray}
\label{jf-fc-crep}
\hat{\varphi}_{k}\Psi &=& \varphi_{k}\Psi, \\
\hat{P}_{\varphi}^{k}\Psi &= &-i \frac{\partial
\Psi}{\partial \varphi_{k}}+i\vert k \vert
\frac{\alpha_{k}^{*} - \beta_{k}}{\alpha_{k}^{*} +
\beta_{k}}\varphi_{-k}\Psi. \label{jf-fc-mrep}
\end{eqnarray}
Notice that, in order to avoid an excessively
complicated notation, we have used the same symbols as
in Eqs. (\ref{jo-fc-crep}) and (\ref{jo-fc-mrep}) to
denote quantum operators and states in the
$j_{t_f}$-SR. For completeness, let us also present
the form of  the annihilation and creation operators
of the $j_{t_f}$-SR, which are given by
\begin{eqnarray}
\hat{\bar{b}}_{k}(t_f)&=&\frac{\alpha_{k}^{*}
+\beta_{k}}{\sqrt{2|k|}}
\frac{\partial}{\partial \varphi_{-k}}, \nonumber \\
\hat{\bar{b}}_{k}^{\dagger}(t_f)&=& -\frac{
\alpha_{k}+\beta^{*}_{k}}{\sqrt{2|k|}}
\frac{\partial}{\partial\varphi_{k}}+
\frac{\sqrt{2|k|}}{\alpha_{k}^{*}+\beta_{k}}
\varphi_{-k}. \label{ann-crea-op-jf}
\end{eqnarray}
As we have discussed above, they represent the
classical variables
$\{(\bar{b}_{k}(t_{f}),\bar{b}^{*}_{k}(t_{f}))\}$. The
quantization of the variables $\{(b_{k},b^{*}_{k})\}$
in this representation can be obtained from (the
inverse of) relations (\ref{q-evolback}), or from Eqs.
(\ref{jf-fc-crep}) and (\ref{jf-fc-mrep}), using
relation (\ref{ann-cre-like}).

Let us now analyze the issue of unitarity in this
context, namely, the unitary equivalence between the
$j_0$-SR and the $j_{t_f}$-SRs. We first remark that,
since unitarity is granted for any finite number of
degrees of freedom, unitary equivalence (for a case of
compact spatial topology such as the present one)
rests just on the behavior of the high frequency
modes. In our case, the asymptotic limit for large $k$
of the sequences $\beta_{k}(t,t_0)$ and
$\alpha_{k}(t,t_0)$ is zero and one, respectively.
Therefore, the factors in the measure
(\ref{measure-b-tf}) and the momentum operators
(\ref{jf-fc-mrep}) approach the corresponding
expressions for the $j_0$-SR. Actually, this is a
necessary condition for unitarity, but not sufficient.
Unitary equivalence between the $j_{t_f}$ and the
$j_0$ representations amounts to requiring that
$j_{t_f}-j_0$ be a Hilbert-Schmidt operator. In turn,
this is equivalent to the summability of the sequences
$\{|\beta_k|^2\}$, a condition which is indeed
satisfied, as shown in Refs. \cite{ccm1,ccm2}. So, all
the representations in the 1-parameter family of
$j_{t_f}$-SRs are equivalent to the $j_0$-SR, and
hence any two members of the family are equivalent to
each other.

Consider now in more detail the momentum operators
(\ref{jf-fc-mrep}), and in particular the extra
multiplicative term (that cannot be obtained from the
measure)
\begin{equation}
-B_{k}^{-1}A_{k}=2\vert k\vert
\frac{{\rm{Im}}(\alpha_{k}\beta_{k})}{\vert
\alpha_{k}^{*}+\beta_{k}\vert^{2}},
\end{equation}
coming from the diagonal component
$2{\rm{Im}}(\alpha_{k}\beta_{k})$ in $(j_{t_f})_k$.
The presence of this term means that the unitary group
generated by the momentum operators is not simply the
natural unitary implementation in
$L^{2}({\cal{T}}^{\star},\d\mu_{t_f})$ of translations
(by elements of $\cal T$) in ${\cal T}^{\star}$. In
addition to the contribution coming from the
transformation under translations of the
quasi-invariant measure $\mu_{t_f}$ [which corresponds
to the term $-iB_k^{-1}$ in Eq. (\ref{jf-fc-mrep})],
the elements of that unitary group carry additional
(nonconstant) phases. Such phases, responsible for the
extra term in Eq. (\ref{jf-fc-mrep}), can be viewed in
our case as generated by the unitary transformation
$T:L^{2}({\cal{T}}^{\star},\d\mu_{t_f})\to
L^{2}({\cal{T}}^{\star},\d\mu_{t_f})$, with
\begin{eqnarray}
(T \Psi)[\varphi_k] & = &\exp\left(i
\sum_{k=1}^{\infty}B_k^{-1}A_k|\varphi_k|^2\right)
\Psi[\varphi_k] \nonumber \\
\label{z6} & = &
\exp\left(-i\sum_{k=1}^{\infty}\frac{2|k|{\rm
Im}(\alpha_k\beta_k)}{|\alpha_k^*+\beta_k|^2}
|\varphi_k|^2\right)\Psi[\varphi_k].
\end{eqnarray}
In fact, one can check that $T^{-1}$ maps the
$j_{t_f}$-SR to the representation defined by the
complex structure ${\overline{j}}_{t_f}$, with
\begin{equation}
\label{z7}
({\overline{j}}_{t_f})_k=\left(\begin{array}{cc} 0 &
-\frac{|\alpha_{k}^{*}+\beta_{k}|^{2}}{|k|} \\
\frac{|k|}{|\alpha_{k}^{*}+\beta_{k}|^{2}} &
0
\end{array} \right).
\end{equation}
It is also worth noting that the summability of
$\{|\beta_k|^2\}$ guarantees that the unitary
transformation $T$ is well defined.\footnote{In
general, the presence of  phases in the unitary
representation  of the group of (appropriate)
translations in the quantum configuration space is a
source of unitary inequivalence in quantum field
theory, in addition to the  existence of nonequivalent
quasi-invariant measures in infinite dimensions (see,
e.g., Refs. \cite{GV,BSZ}). From the viewpoint of the
momentum operators, rather than from that of the
corresponding unitary group, this issue was addressed
more recently in Ref. \cite{ccq-cqg}, where the
possible lack of unitary equivalence between
representations with and without an extra linear term
in the momentum operators was discussed, and related
to the possibility or impossibility of defining
unitarity transformations of the type (\ref{z6}).}

Adopting the above perspective, the unitary
transformation mapping the $j_{t_f}$-SR to the
$j_0$-SR can be obtained as the composition of
$T^{-1}$ with the natural unitary transformation
between the ${\overline{j}}_{t_f}$-SR and the
$j_0$-SR, namely $\Psi\mapsto \left(\d\mu_{t_f}/\d\mu
\right)^{1/2}\Psi$. We also notice that the existence
of both derivatives $\d\mu_{t_f}/\d\mu$ and
$\d\mu/\d\mu_{t_f}$, i.e. the mutual absolute continuity
of the Gaussian measures, depends on whether the
operator $C_{t_f}C^{-1}-{\bf 1}$ is Hilbert-Schmidt,
where $C$ and $C_{t_f}$ denote the covariances of
$\mu$ and $\mu_{t_f}$, respectively. In the present
case this leads to the condition that
$\{|\alpha_k^*+\beta_k|^2-1\}$ be a square summable
sequence. Again, this is ensured by the summability of
$\{|\beta_k|^2\}$.

Summarizing, the unitary transformation mapping the
$j_{t_f}$-SR to the $j_0$-SR is the multiplicative
transformation
\begin{equation}
\label{z8} L^{2}({\cal{T}}^{\star},\d\mu_{t_f})\ni
\Psi\mapsto \left(\frac{\d\mu_{t_f}}{\d\mu}
\right)^{1/2}\exp\left(i\sum_{k=1}^{\infty}
\frac{2|k|{\rm Im}(\alpha_k\beta_k)}{
|\alpha_k^*+\beta_k|^2}|\varphi_k|^2\right)\Psi\in
L^{2}({\cal{T}}^{\star},\d\mu).
\end{equation}
Of course, the multiplicative factor in this
expression is simply the image of the unit functional
$\Psi_{0}[\varphi_{k}]=1$ of the $j_{t_f}$-SR, and
therefore supplies the state $|0,t_f\rangle$
(\ref{vac-rel}) in the $j_0$-SR, namely, it coincides
with $\Psi_0^{(t_f)}$ given in Eq. (\ref{jf-vs}) [one
can check this by introducing the explicit form of
$\d\mu_{t_f}/\d\mu$ obtained from Eqs. (\ref{measure-b})
and (\ref{measure-b-tf})].

Finally, we want to comment that any unitary
transformation between two SRs admits a form like that
displayed in Eq. (\ref{z8}). In fact, given two
normalized measures $\mu_{1}$ and $\mu_{2}$ (not
necessarily Gaussian), if a unitary transformation $U$
exists such that it maps one SR to the other, then it
is necessarily of the multiplicative form $\Psi\mapsto
\Psi_0\Psi$, where $\Psi_0$ is the image under $U$ of
the unit functional.\footnote{This can be seen using
the fact that, by construction, configuration
operators such as the unitary groups generated by the
basic field operators generate a dense set when
applied to the unit functional. On the other hand, the
action of the configuration operators is the same in
both representations. Thus, for
$\Psi=e^{i\bar{\varphi}(f)}$,
$U\Psi=Ue^{i\hat{\bar{\varphi\,}}[f]}
1=Ue^{i\hat{\bar{\varphi}\,}[f]}U^{-1}
\Psi_0=e^{i\bar{\varphi}(f)}\Psi_0$. The general
expression follows from linearity and continuity.}
Moreover, the identity $\int|\Psi|^2
\d\mu_{1}=\int|\Psi_0|^2|\Psi|^2 \d\mu_{2}$, valid
$\forall\Psi$, implies that $\mu_{1}$ is continuous
with respect to $\mu_{2}$, with $\d\mu_{1}/\d\mu_{2}=
|\Psi_0|^2$. By interchanging the roles of $\mu_{1}$
and $\mu_{2}$, one concludes that the measures are
mutually continuous. Thus, the equivalence of the
measures is a necessary condition for the unitary
equivalence between two SRs, and any possible unitary
equivalence is of the form $\Psi\mapsto
(\d\mu_{1}/\d\mu_{2})^{1/2}e^{iF}\Psi$, where $F$ is a
real functional. As one can easily realize from the
discussion of Ref. \cite{ccq-cqg}, in the case of two
representations defined by equivalent complex
structures, the functional $F$ is a bilinear form of
the type appearing in Eq.~(\ref{z6}) and its
introduction results in a modification of the action
of the momentum operators by linear terms.

\section{Conclusion}
\label{sec:6}

In full canonical quantum gravity formulated on a
compact spatial section $\Sigma$, there is no
fundamental notion of time. There is no Hamiltonian,
and therefore no time with respect to which one might
define evolution (this is one of the manifestations of
the notorious problem of time). The Gowdy model that
we have considered here is somewhat special in this
respect since, through a partial gauge fixing, a
particular notion of internal time is introduced in
order to ``de-parametrize'' the theory. Even when this
parameter has no physical meaning in the final
description, it is used as an intermediate step in
order to construct the corresponding physical operators that
define the true quantum geometry. This is the strategy that
has also been followed in the quantization of homogeneous
cosmologies \cite{misner1}.
Therefore, within
the model, it  is important to implement this notion
of time evolution in a unitary way. Furthermore, the
strategy that we have followed of implementing the internal
notion of time at the quantum level, together with the remaining
gauge group, receives  support from the fact that a
quantization with such properties exists
\cite{ccm1,ccm2} and is essentially unique
\cite{cmv,ccmv}. This consistent quantization has to
be contrasted to a previous
proposal \cite{pierri}
that does not admit an unitary time evolution \cite{non-uni}.

The purpose of this paper was to bridge the gap
between the formalism of Refs. \cite{ccm1,ccm2} and
the standard formulation of canonical quantum gravity,
and thus to recast the quantization of the Gowdy model
into the Schr\"odinger functional representation,
where the states of the theory are functionals
$\Psi(\overline{\varphi})$ on the quantum
configuration space. Let us now summarize the results
found here. First, we have constructed the
Schr\"{o}dinger functional version of this quantum
Gowdy model, and analyzed the (unitary) time evolution
in this context. Second, we have solved the remaining
constraint that is present in the model. In this way,
we have been able to define the space of physical
states in the Sch\"odinger picture, where unitary
evolution is again well defined.

As a general strategy, we have approached the problem from
a functional perspective. In this fashion, we have constructed
explicitly the 1-parameter family of representations
that gives rise to the quantum description at any
time. These different representations are unitarily equivalent
precisely because time evolution is unitarily
implementable. We have discussed in some detail the
unitary transformations between such representations,
confirming that, in the Schr\"odinger representation,
they are associated with a
corresponding 1-parameter family of mutually
continuous measures in the quantum configuration
space.
This has to be contrasted with the functional description \cite{torre-sr}
of the quantization proposed
in Ref. \cite{pierri}, which does not admit unitary evolution.
In that case, the fact
that the dynamics fails to be unitarily implementable  implies that
any two representations at different times correspond to
inequivalent measures. In fact, a 1-parameter
family of mutually singular measures is obtained in that case \cite{torre-sr}.

To conclude, our functional representation
leads to a consistent framework where the standard
probabilistic interpretation of quantum physics is
applicable. In particular, the Heisenberg and
Schr\"{o}dinger pictures are well defined and
conciliated.
The present description  can thus be taken as a starting
point
for a detailed study of the quantum geometric aspects
of linearly polarized Gowdy models.

\section*{Acknowledgements}

This work was supported by the Spanish MEC Projects
FIS2005-05736-C03-02 and FIS2006-26387-E/, the CONACyT
U47857-F grant, the Joint CSIC/CONACyT Project
2005MX0022, the Portuguese FCT Project
POCTI/FIS/57547/2004, the NSF PHY04-56913 grant and
the Eberly Research Funds of Penn State.

\end{document}